\documentclass[a4paper]{mem}
\usepackage{natbib}
\usepackage{graphicx}
\usepackage[a4paper]{hyperref}
\idline{73}{23}
\begin{document}
   \title{TNG photometry of the open cluster NGC\,6939
}

   \author{G. Andreuzzi \inst{1},
   A. Bragaglia \inst{2},\\
   M. Tosi \inst{2}
          \and
          G. Marconi \inst{3}
\thanks{Based on observations obtained at the Telescopio Nazionale Galileo}
}
   \offprints{G. Andreuzzi}
\mail{gloria@mporzio.astro.it}
   \institute{\
  INAF--Osservatorio Astronomico di Roma,
   via Frascati 33, 00040 Monteporzio Catone Roma (Italy)
     \email{gloria@mporzio.astro.it}\\
   \and  INAF--Osservatorio Astronomico di Bologna,
   Via Ranzani 1, I-40127 Bologna (Italy)
   \email{angela, tosi@bo.astro.it} \\
    \and ESO, Alonso de Cordova 3107, Vitacura, Santiago (Chile)
 }

   \abstract{

We present CCD UBVI photometry of the intermediate age open cluster NGC 6939
for three TNG-DOLORES fields. The fields A and B cover the center of the
cluster; the third one is located about 30$\arcmin$ away, and is used for field
stars decontamination.

The V-I, B-V and U-B color-magnitude diagrams (CMDs), obtained joining very 
different exposures show: i) a Main Sequence (MS) extending down to V = 24,
 much deeper ($\sim$ 5 magnitudes) than any previous study; ii) a clearly
defined Turn Off (TO) and iii) a well populated Red Giant Clump (RC) at about
V = 13.

   \keywords{Hertzsprung-Russell (HR) diagram --
     open clusters and associations: general --
     open clusters and associations: individual: NGC 6939
               }
   }
   \authorrunning{G. Andreuzzi et al.}
   \titlerunning{TNG photometry of NGC\,6939}
   \maketitle

\begin{figure}
   \centering
   \includegraphics[width=10cm, height=8cm]{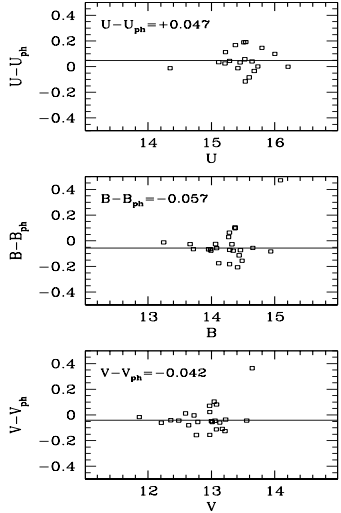}
      \caption{Comparison of our U,B,V photometry with the photoelectric
        values given in Mermilliod et al (1994).      
              }
         \label{fig-photo}
\end{figure}

\section{Introduction}

Open clusters are ideal tracers of the Galactic disk properties, covering
a large interval in position, metallicity, and age. They can be used to study
not only the present day situation, but also the time evolution of the disk
(e.g. Friel 1995).
We are building a large sample of open clusters homogeneously studied
(see Bragaglia \& Tosi 2003 and references therein), concentrating on the
old ones. In this paper we present our results on another open cluster.

Given its relative proximity, NGC 6939 (RA(2000) = 20:31:32, DEC(2000) =
+60:39:00, or l = 95.88,  b = 12.30) has been the target of
several studies in the past: the first bibliographic entry is 80 years ago
(Kustner 1923) but, surprisingly, the first CCD data appeared only in 2002
(Rosvick \& Balam 2002, hereafter RB02). As
usual, the cluster parameters found in literature do not agree well, and we
present new and improved photometric results to be used in the future
to give new determinations for this intermediate age open cluster.

Photometry has been previously presented by several authors, but old
photographic photometry only reached about one magnitude below the main
sequence Turn-Off.  Mermilliod, Huestamendia, $\&$ del Rio (1994) took UBV
photoelectric photometry of 37 members stars  all in the red
clump phase, with the intent of discriminating between different evolutionary
models (with or without overshooting) by comparison with isochrones. The recent
work by RB02 has presented the first deep BVI CCD data.
They used the 1.85m Dominion Astrophysical Observatory telescope, covering the
central part of NGC 6939 (more or less our field A, see later). Their CMDs show
considerable scatter, which they do not attribute to contamination from field
stars (even if they could not prune their diagrams on the MS, since
no proper motion survey on this cluster has reached so deep), but to
differential reddening [E(B-V) = 0.29 to 0.41], which also influences the
distance derivation [$(m-M)_V$ = 12.21 to 12.39]. Using the Girardi et al.
(2000) solar metallicity isochrone, they obtain a cluster age of 1.6 $\pm$ 0.3
Gyr

\section {Observations and data reduction}

\begin{figure*}
   \centering
   \includegraphics[height=8.5cm, width=12cm]{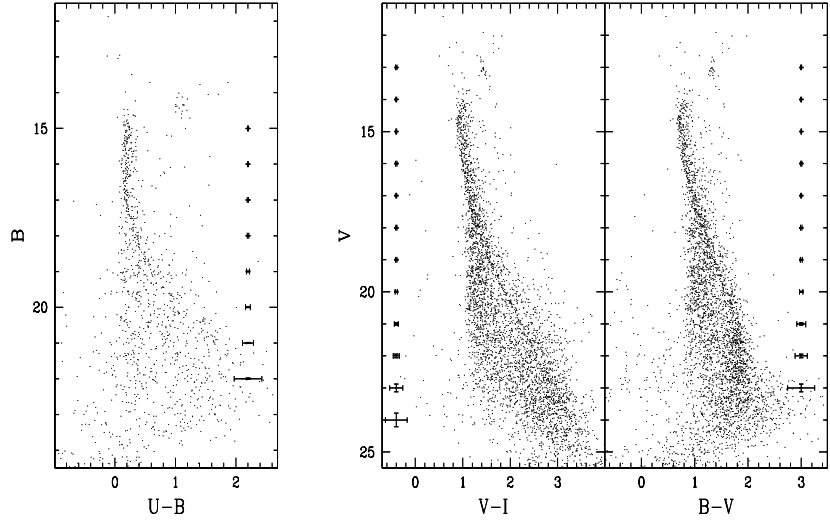}
      \caption{V-I, B-V CMDs for NGC\,6939.
           The mean errors per interval in magnitude V are also plotted
}
         \label{fig-cmdc1}
\end{figure*}

\begin{figure*}
  \centering
   \includegraphics[height=8.5cm, width=12cm]{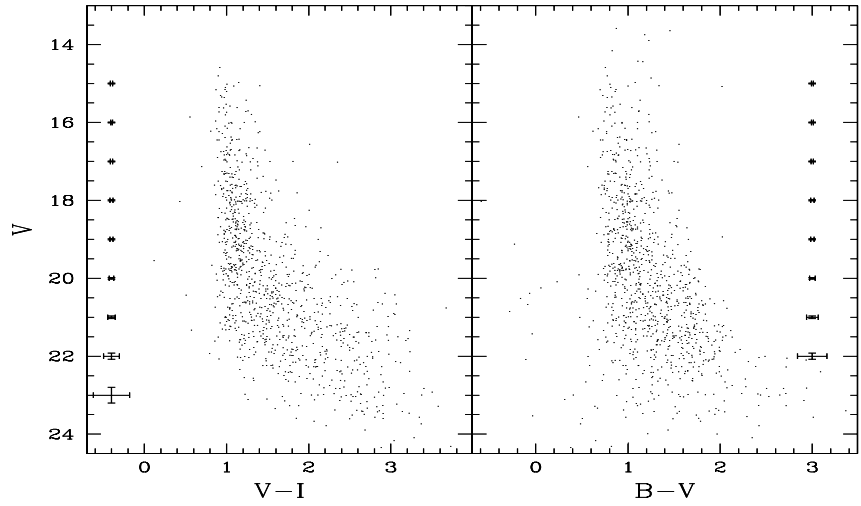}
      \caption{V-I, B-V CMDs for the external field.
              The mean errors per interval in magnitude V are also plotted.
              }
         \label{fig-cmdf}
\end{figure*}

Our data were acquired at the Telescopio Nazionale Galileo,
using  DOLORES (Device  Optimized for the LOw RESolution)
on two different runs (see Table 1 for details).
In both cases the images have been observed with a scale of 0.275
$\arcsec$/pix, (field of view 9.4$\arcmin$ $\times$ 9.4$\arcmin$).
Three different fields have been observed: one
centred on the cluster (field A), one slightly North of it (field B), and the
last (that may be used for field stars decontamination) about 30$\arcmin$ away.

\begin{table*}
\begin{center}
\caption{Log of our observations. For each field we list the dates of the
observations, the filters used and the corresponding ranges of
exposure time (in seconds)}
\vspace{5mm}
\begin{tabular}{lclllll}
\hline\hline
Field & coords (2000) & UT dates & t$_U$ & t$_B$ & t$_V$ & t$_I$\\
\hline
Field A & 20:31:31 +60:29:21 & 2000 Nov 24, 25 & 60-1200 & 600-20 & 300-60 & 300-10\\
        &                    & 2001 Aug 17     &         & 5-2    & 10-2   & 5-2  \\
Field B & 20:31:30 +60:46:51 & 2000 Nov 26     &         & 600-40 & 300-20 & 300-20\\
        &                    & 2001 Aug 18     &         & 5-2    & 5-2    & 5-2\\
external & 20:31:32 +60:09:23 & 2000 Nov 25    &         & 600-20 & 300-10 & 300-20\\
\hline
\end{tabular}
\end{center}
\label{oss}
\end{table*}

Corrections to the raw data for bias, dark and flat-fielding
were performed using the standard IRAF routines.
Subsequent data reduction and analysis was done following the same procedure
for the three data-sets and using the Daophot-Allframe packages
(Stetson-version 3, 1997) with a quadratically varying point spread
function. The identified candidates were measured on each of
the individual I, V, B and U frames of each field and two final catalogs
(one for the cluster and one for the comparison field) have been created.

The calibration equations were derived by using the objects in the
standard areas PG0231+051 and Rubin 149, plus the two isolated stars G156--31,
and G26--7 (Landolt 1992):

$$  U = u +0.1796 \times (u-b) -0.9120   $$ 
$$  B = b +0.0525 \times (b-v) +1.4187   $$ 
$$  V = v -0.1490 \times (b-v) +1.2389   $$ 
$$  I = i +0.0282 \times (v-i) +0.7854   $$ 

where u, b, v, i, are the aperture corrected instrumental magnitudes, after
further correction for extinction
and for exposure time, and U, B, V, I are the
output magnitudes, calibrated to the Johnson-Cousins standard system.

To check our calibration we directly compared our U, B, V  magnitudes for  24
stars in common with the photoelectric values given in Mermilliod et al.
(1994). This study has shown that no trend is present between the two data-sets
but only zero point shifts, as indicated in Fig. \ref{fig-photo}.

\section{Data analysis and discussion}

The final cluster CMDs are shown in Fig. \ref{fig-cmdc1}, 
where we also plot the mean photometric errors per interval of magnitude, to
understand the data quality.

We emphasize that these CMDs represent by far the best photometric results
found in literature for this cluster.  
From the figure it is possible to see a very clear MS extending down to
V $\sim$ 24, nearly 5 mag deeper than the only
previous CCD photometry (RB02), together with
a prominent red giant clump, including $\sim$ 40 stars.

A contamination from field stars is also evident, especially
in the fainter parts, while the brighter stars are almost free from field
interlopers. In order to account for field stars contamination, we may use
data from the external field, observed far enough from the cluster so
that the contamination by cluster members - if any - is minimal.

Figure \ref{fig-cmdf} shows the resulting V-I and B-V CMDs for the stars
located in this comparison field.

In the following part of our study,  the observed CMDs will be compared to
synthetic ones to  derive the cluster parameters distance, reddening, age and
approximate  metallicity, using a procedure amply described in Tosi et al.
(1991) and in  the other papers of the open clusters series (see e.g. Bragaglia
\& Tosi 2003).
These synthetic CMDs are built using different evolutionary tracks with the
same number of stars, error distribution, and incompleteness factors of
the observed ones, taking also into account the possibility of binary
systems.

\begin{acknowledgements}
Financial support to this project has come from the MURST-MIUR through
Cofin98 ''Stellar Evolution'', and Cofin00 ''Stellar Observables of Cosmological
Relevance''.
We warmly thank M. Bellazzini for the data acquired on
August 2001.
\end{acknowledgements}

\bibliographystyle{aa}

\begin{thebibliography}{}

\bibitem[{}]{}
Bragaglia A., Tosi M. 2003, MNRAS, in press (astro-ph/0303662)
\bibitem[{}]{}
Friel E.D. 1995 ARA\&A, 33, 38
\bibitem[{}]{}
Girardi, L., Mermilliod, J.-C., Carraro, G. 2000 A\&A, 354, 892
\bibitem[{}]{}
Kustner, F. 1923, Veroeffentlichungen der Universitaets-Sternwarte zu Bonn, 18
\bibitem[{}]{}
Landolt, A., U. 1992, AJ 104(1), 340
\bibitem[{}]{}
Mermilliod, J., C., Huestamendia, G., del Rio, G. 1994 A\&AS, 106, 419
\bibitem[{}]{}
Rosvick, J., M., Balam, D. 2002, AJ 124, 2093
\bibitem[{}]{}
Tosi, M., Greggio, L., Marconi, G., Focardi, P. 1991, AJ 102, 951


\end{thebibliography}

\end{document}